# A Constructive Procedure for Modeling Categorical Variables: Log-linear and Logit Models

Philip E.Cheng[1], Jiun-Wei Liou[1], Hung-Wen Kao[2], Michelle Liou[1,3*]

[1]Institute of Statistical Science, Academia Sinica, Taipei 115, TAIWAN

[2]Department of Radiology, Tri-Service General Hospital, National Defense Medical Center, Taipei 114, TAIWAN

[3]Imaging Research Center, Taipei Medical University, Taipei 110, TAIWAN

## Abstract

Association between categorical variables in contingency tables is analyzed using the information identities based on multivariate multinomial distributions. A scheme of geometric decompositions of the information identities is developed to identify indispensable predictors and interaction effects in the construction of concise log-linear and logit models; it suggests a new approach for selecting parsimonious log-linear and logit models which would facilitate the search for the minimum AIC models as a byproduct. The proposed constructive schemes are illustrated along with the analysis of a contingency data table collected in a study on the risk factors of ischemic cerebral stroke.

**Correspondence:** Michelle Liou, Institute of Statistical Science, Academia Sinica, Taipei, Taiwan, 115. Email: mliou@stat.sinica.edu.tw.

# 1. Introduction

Log-linear and logit regression models have been widely used for statistical inference of discrete variables in the family of generalized linear models (GLMs) (Bishop, Fienberg & Holland, 1975; McCullagh and Nelder, 1989; Everitt, 1992; Agresti, 2002). Analogous to the ANOVA decomposition, a log-linear model defines the logarithms of expected cell counts of categorical variables in a contingency table as a linear regression equation of marginal and interaction effects. Being the counterpart of log-linear models, a logit model predicts the odds of a binary variable using all the other variables in the table as predictors. It is desirable to acquire parsimonious log-linear and logit models with concise data interpretation in application. Among criteria of model selection, the AIC is particularly designed for achieving optimal prediction accuracy, which is comparable to cross validation (Akaike, 1973; Stone, 1974). A list of commonly used software packages includes the AIC, $AIC_c$, BIC and the standard stepwise forward/backward selection procedures (Sugiura, 1978; Schwarz, 1978). Discussions of these procedures with applications to various data examples can be found in Burnham and Anderson (2002), Fahrmeir and Tutz (1994), etc. There were also review studies assessing stepwise subset selection methods in linear regression including the AIC and BIC criteria, compared with Bayesian model averaging, penalized regression and elastic net methods (Morozova et al., 2015; Walter and Tiemeier, 2009).

  The notion of parsimony has often been regarded as using models with purely main effects in fields of applied sciences, e.g., the Rasch model in educational statistics (cf. Fischer and Molenaar, 1995). Parsimonious log-linear and logit models are recommended to offer simple data interpretation, but not confined to those with only main effects, because interaction effects among the explanatory variables can be natural and indispensable elements in a model. For instance, drug-drug interactions and effects of comorbidity on drug efficacies are often considered in medical data analysis (Cleophas et al., 2007). Gene-gene and gene-environment interaction effects on diseases are also recognized as essential in the analysis of gene expression data (Castaldi et al., 2017). Testing and interpretation of interaction effects in these models inevitably affects proper model selection and must be fully addressed (Jaccard, 2001), but almost all the existing model selection procedures have not discussed this aspect. In this study, we will propose an information theoretical approach to building log-linear and logit models through identifying the indispensable interaction effects after the main-effect variables are selected.

A recent study showed that the analysis of mutual information (MI) presents a geometric interpretation of the association between categorical variables based on the invariant Pythagorean laws of testing for odds ratios in 2-way tables (Cheng et al., 2008). Extensions of this invariant law to multi-way tables further characterized the geometry for testing conditional independence between two variables given the other covariates as the hypotenuse of a right triangle whose two legs define independent tests for the interaction and the partial association, respectively (Cheng et al., 2010). In multivariate tables, the MIs of random vectors are decomposed as sums of lower-dimensional MI terms and conditional mutual information (CMI) terms which are orthogonal in information. Different partitions of the MI and CMI terms of a table yield various forms of information identities, which characterize various associative relations of the variables, while the total information shared by these identities are equal (Cheng et al., 2007). Specifically, decompositions of the MI of associative variables may be arranged to characterize the insignificant interaction effects such that the least significant main-effect predictors and their interaction effects can be identified and selected into the model. This is directly applicable to both log-linear modeling and logit modeling given a contingency data table.

To develop an information theoretical approach to constructing log-linear and logit models using indispensable main and interaction effects, the study is laid out as follows. In Section 2, a review is devoted to the basic elements of statistical information theory. The MI, CMI terms in two- and three-way tables are defined, along with the relations to main and interaction effects in specific log-linear models. Basic information identity in a 3-way table is illustrated with extensions to multi-way tables. In Section 3, a dataset collected in a clinical study on the risk factors of ischemic stroke is applied to illustrate the method and theory given in Section 2. The focus is placed on the information approach to constructing log-linear models by testing interaction effects. A sequence of testing for the least indispensable interaction effects is carried out such that the most parsimonious log-linear models (almost information-equivalent) are identified. In Section 4, the results of log-linear models are applied to facilitate the analysis of parsimonious logit models. As illustrated with the same dataset, the MI approach to logit modeling is initiated with identifying the indispensable predictors for the binary response target, following the CMI analysis for the log-linear model. Given the predictors, the tested interaction effects in the log-linear models can be directly used to identify the required interaction parameters of a logit model. These analyses furnish the entire scheme of constructing both log-linear and logit models based the MI analysis of information identities. As a byproduct, the minimum AIC logit model using the same predictors can be easily identified in a neighborhood of the acquired MI logit model. We conclude the study with a brief discussion of potential extensions of the geometric information analysis to GLMs involving categorical and continuous variables.

## 2. Log-linear Models and Information Identities

The classical studies of partitioning chi-squares in a three-way table (Lancaster, 1951; Mood, 1950; Snedecor, 1958) inspired the development of the log-linear model (Goodman, 1964; Kullback, 1968), which was subsequently used for decades to measure associations among categorical variables (Birch, 1964; Bishop et al., 2007; Goodman and Kruskal, 1979). In this section, we demonstrate the link between a few basic log-linear models and their corresponding information identities in three- and multi-way tables.

Let (X, Y, Z) denote a three-way $I \times J \times K$ contingency table with the joint probability density function (pdf) $f_{X,Y,Z}(i,j,k)$, for $i = 1, \ldots, I$, $j = 1, \ldots, J$ and $k = 1, \ldots, K$. The Shannon entropy defines the basic equation in terms of joint and marginal probabilities:

$$H(X) + H(Y) + H(Z) = I(X; Y; Z) + H(X, Y, Z), \tag{1}$$

where

$$H(X, Y, Z) = -\sum_{i,j,k} f_{X,Y,Z}(i,j,k) \cdot \log[f_{X,Y,Z}(i,j,k)]$$

is the joint entropy,

$$H(X) = -\sum_i f_X(i) \cdot \log[f_X(i)]$$

is the marginal entropy of X, and

$$I(X; Y; Z) = \sum_{i,j,k} f_{X,Y,Z}(i,j,k) \cdot \log\left\{\frac{f_{X,Y,Z}(i,j,k)}{f_X(i) f_Y(j) f_Z(k)}\right\}$$

denotes the MI between X, Y and Z (Cover and Thomas, 2006; Kullback and Leibler, 1951). There exists a geometric aspect of the MI, which defines the Kullback-Leibler divergence from the joint pdf to the space of products of marginal pdfs, that is, the space of the null hypothesis of independence (Cheng et al., 2006; Cheng et al., 2008). By factoring the joint log-likelihood, an orthogonal partition of the MI among the three variables can be expressed as the following information identity

$$I(X; Y; Z) = I(X; Z) + I(Y; Z) + I(X; Y|Z). \qquad (2)$$

The right-hand side of (2) admits three equivalent identities by exchanging the common variable Z with either X or Y. Here, a two-way MI term such as $I(X; Z)$ is defined with the marginal (X, Z) table using an analog of the three-way table in (1). The conditional mutual information (CMI) $I(X; Y|Z)$ on the right-hand side of (2) defines the expectation of the log-likelihood ratio for testing the conditional independence between X and Y across levels of Z. Based on the multivariate multinomial likelihood, Equation (2) and its sample version are valid with the same formula, that is, the same equation holds when the MIs and CMIs are replaced by their sample analogs. In practice, the sample version of (2) can be expressed as

$$\begin{aligned}
\hat{I}(X; Y; Z) &= 2N \sum_{i,j,k} \hat{f}_{X,Y,Z}(i,j,k) \log\left(\frac{\hat{f}_{X,Y,Z}(i,j,k)}{\hat{f}_X(i)\hat{f}_Y(j)\hat{f}_Z(k)}\right) \\
&= 2N \sum_{ik} \hat{f}_{XZ}(i,k) \log\left(\frac{\hat{f}_{XZ}(i,k)}{\hat{f}_X(i)\hat{f}_Z(k)}\right) + 2N \sum_{jk} \hat{f}_{YZ}(j,k) \log\left(\frac{\hat{f}_{YZ}(j,k)}{\hat{f}_Y(j)\hat{f}_Z(k)}\right) \\
&\quad + 2N \sum_k \left[\sum_{ij} \hat{f}_{XY|Z}(i,j|k) \log\left(\frac{\hat{f}_{XY|Z}(i,j|k)}{\hat{f}_{X|Z}(i|k)\hat{f}_{Y|Z}(j|k)}\right)\right] \\
&= \hat{I}(X; Z) + \hat{I}(Y; Z) + \hat{I}(X; Y|Z), \qquad (3)
\end{aligned}$$

where N is the total sample size, and the constant 2N is used for the approximations to suitable chi-square distributions. The notation $\hat{f}_{XZ}(i,k)$ denotes the estimated joint pdf in the (i, k) cell, and $\hat{f}_X(i)\hat{f}_Z(k)$ is the estimated product pdf under the assumption of independence. Other notations in (3) are defined by analogy. It follows that $\hat{I}(X; Z)$, $\hat{I}(Y; Z)$ and $\hat{I}(X; Y|Z)$ on the right-hand side of (3) are asymptotically chi-square distributed with (I-1)(K-1), (J-1)(K-1) and (I-1)(J-1)K degrees of freedom (*df*s), respectively (Cheng et al., 2007; Kullback, 1968).

In application, the sample MI, denoted by $\hat{I}$(X; Y), is the LR deviance statistic for testing for two-way independence between X and Y, which is the same test for the hypothesized log-linear model, denoted by {X, Y}, composed of the intercept plus the two main effects X and Y. The hypothesis of conditional independence in a three-way table defines the null CMI, *I*(X; Y| Z) = 0, which also defines the hierarchical log-linear model {XZ, YZ}. The three-way Pythagorean law (P-law) depicts that *I*(X; Y| Z) is the hypotenuse of a right triangle with two orthogonal legs: the interaction effect {XYZ}, denoted by *Int*(X; Y; Z), measuring the heterogeneous association between X and Y across the levels of Z, and the partial association, denoted by *Par*(X; Y|Z), measuring the homogeneous association between X and Y across the levels of Z. Specifically, the CMI term on the right-hand side of (2) is expressed as the sum of two orthogonal components:

$$I(X; Y|Z) = Int(X; Y; Z) + Par(X; Y|Z). \qquad (4)$$

Similar to (2) and (3), the sample analogs of these terms in (4) also satisfy the identity in (4). The sample CMI $\hat{I}$(X; Y|Z) is the MLE of conditional independence, which is the last summand in (3). The MLE $\widehat{Int}$(X; Y; Z) in (4) can be computed using the iterative proportional fitting or the Newton-Raphson procedure (Bishop et al., 2007), and the MLE of the partial association $\widehat{Par}$(X; Y|Z) can be obtained by the difference between $\hat{I}$(X; Y|Z) and $\widehat{Int}$(X; Y; Z). The Pythagorean law in (4) characterizes *I*(X; Y|Z) to be the hypotenuse of a right triangle with two legs: the interaction *Int*(X; Y; Z) and the uniform association *Par*(X; Y|Z). It was proved that the LR statistic $\hat{I}$(X; Y|Z), testing conditional independence using the chi-square distribution with (I-1)(J-1)K *df*s, can be decomposed

into a two-step LR test. The first step directly tests the hypothesis of no interaction between X and Y across the levels of Z using $\widehat{Int}$(X; Y; Z) with (I-1)(J-1)(K-1) *df*s, and, only if this hypothesis is accepted, the hypothesis of uniform association is tested using $\widehat{Par}$(X; Y|Z) with (I-1)(J-1) *df*s. In applications, it implies that, given a significant test for the said conditional independence at the usual level 0.05, the hypothesis of no interaction is legitimately tested against a smaller level than 0.05 (cf. Figure 1 in Cheng et al., 2010).

The extension of (2) to multi-way tables leads to information identities in general cases. For instance, the association between a variable T and three predictors X, Y and Z can be measured using the following mutual information (MI) identity:

$$I(T; X; Y; Z) = I(X; Y; Z) + I(\{X, Y, Z\}; T)$$
$$= I(X; Z) + I(Y; Z) + I(X; Y|Z)$$
$$+ [I(T; Z) + I(T; Y|Z) + I(T; X|\{Y, Z\})]. \quad (5)$$

By using an additional variable T in (3), the term $I(\{X, Y, Z\}; T)$ in (5) is used to measure the association between T and $\{X, Y, Z\}$. If T denotes the target in a logit model, the terms in the brackets of (5) would describe the regression of T on X, Y and Z. For instance, if the null hypothesis CMI $I(T; X|\{Y, Z\}) = 0$ is retained, then X is dispensable while Y and Z are already in the model. With a vector variable Z, equation (5) allows the prediction of T by more than three variables. Equivalent and useful MI identities to (5) can be obtained by interchanging X and Y, or X and Z (Cheng et al., 2007).

### 3. Geometric Log-linear Modeling

Standard analysis of log-linear models for a contingency table commonly begins with testing model-data-fit of all two-way interactions among the variables, followed by adding

three- and higher-way interactions as needed for fit (Agresti, 2002; Bishop et al., 2007; Christensen, 2006; Everitt, 1992; Whittaker, 2009). In this section, the geometric MI approach is proposed for constructing log-linear models in a multi-way table as an alternative to the conventional analysis. The method is based on testing and deleting dispensable higher-order CMI and interaction terms from an information identity of the full saturated model, and processed step-by-step using the two-step LR tests. In principle, it develops a scheme of constructing parsimonious log-linear models through identifying significant two- and high-way interactions. This will be illustrated using the clinical dataset on the risk factors of ischemic stroke.

In the dataset, the brain computed tomography scans were available from 354 patients diagnosed with ischemic stroke in the middle cerebral arterial (MCA) territory and 1,518 control subjects (Kao et al., 2015). The data were collected during 2006 through 2008 to investigate the association between ischemic stroke and its risk factors using a logit model, and the calcification burden in the MCA territory was of main interest. The target response was the status of stroke patients versus controls (S; 1 = case; 0 = control), and the risk factors consisted of the calcification burden (C; 1 = yes, 0 = no) in the MCA territory, age (A; $1 \geq 60$; $0 < 60$), gender (G; 1 = male, 0 = female), hypertension (H; 1= SBP > 140mm Hg or DBP > 90 mm Hg; 0 = none), diabetes mellitus (D; 1= fasting serum glucose level > 7.8 mmol/L, 0 = otherwise), and smoking (M; 1= smoking over 1 cigarette/day, 0 = none). We first assess the parsimonious log-linear model in the 7-way contingency table.

We proceed by inspecting as many dispensable higher-order interactions among the variables as possible, and begin with identifying the first risk factor giving the least significant (or most insignificant) high-way interactions with other factors. After deleting insignificant higher-way CMI terms, keeping the significant lower-order ones, the first factor will be excluded in the subsequent analysis. The scheme proceeds by inspecting the next insignificant higher-way CMI effects among the remaining factors, and stops at the stage when all lower-order interactions are significant.

The CMI statistics among the 7 factors indicate the risk factor C "calcification burden" giving the least significant association with the other 6 factors. By putting aside the factor C, the next risk factor having this property is M, followed by the factor G. The remaining four factors {S, A, D, H} are found highly associated with each other. Similar to (5), the basic information identity among the seven risk factors can be expressed as

$$\hat{I}(C, M, G, S, D, H, A)$$
$$= \hat{I}(\{A, D, G, H, M, S\}; C) + \hat{I}(\{A, D, G, H, S\}; M) +$$
$$\hat{I}(\{S, A, D, H\}; G) + \hat{I}(\{D, A, H\}; S) + \hat{I}(D; A; H). \quad (6)$$

Information analysis of the terms in (6) will be illustrated in order. The first summand on the right-hand side of (6), the MI between C and other factors, is found to yield four insignificant CMI terms which are marked with asterisks in (7) below,

$$\hat{I}(\{A, D, G, H, M, S\}; C) = \hat{I}(C; M|\{A, D, G, H, S\})^* \ (= 15.232, df = 32, p = .995)$$
$$+ \hat{I}(C; G|\{S, H, D, A\})^* \ (= 9.768, df = 16, p = .878)$$
$$+ \hat{I}(C; D|\{S, H, A\})^* \ (= 5.623, df = 8, p = .689)$$
$$+ \hat{I}(C; H|\{S, A\})^* \ (= 5.057, df = 4, p = .281)$$
$$+ \hat{I}(C; A|S) \ (= 31.449, df = 2, p < 0.001)$$
$$+ \hat{I}(C; S) \ (= 96.972, df = 1, p < 0.001). \quad (7)$$

It is seen that $\widehat{Int}(C; A; S)$ (= 8.234, df = 1, p = .004) offers a significant component of $\hat{I}(C; A|S)$ according to the P-law in (4). Excluding the factor C, the factor M of the second summand in (6) emerges with three insignificant CMIs, that is,

$$\hat{I}(\{A, D, G, H, S\}; M) = \hat{I}(M; A|\{D, G, H, S\})^* \ (= 25.325, df = 16, p = .064)$$
$$+ \hat{I}(M; D|\{G, H, S\})^* \ (= 12.589, df = 8, p = .127)$$
$$+ \hat{I}(M; H|\{G, S\})^* \ (= 5.196, df = 4, p = .268)$$
$$+ \hat{I}(M; S|G\}) \ (= 16.935, df = 2, p < 0.001)$$
$$+ \hat{I}(M; G) \ (=314.210, df = 1, p < 0.001). \qquad (8)$$

It is found that $\widehat{Int}(M; G; S)$ (= 7.224, $df = 1$, $p = .007$) also yields a significant interaction component in the 3-way CMI term in (8). Next, the factor G "gender" of the third summand in (6) is found to yield two insignificant CMIs, that is,

$$\hat{I}(\{S, A, D, H\}; G) = \hat{I}(G; S|\{A, D, H\})^* \ (=11.388, df = 8, p = .181)$$
$$+ \hat{I}(G; H|\{A, D\})^* \ (= 8.695, df = 4, p = .069)$$
$$+ \hat{I}(G; D|A\}) \ (= 18.891, df = 2, p < 0.001)$$
$$+ \hat{I}(G; A) \ (= 13.714, df = 1, p < 0.001). \qquad (9)$$

As in (8), $\widehat{Int}(G; D; A)$ (= 13.529, $df = 1$, $p < 0.001$) is a significant interaction component of the third term in (9). With the factor S "stroke status" of the fourth term in (6), the analysis is essentially completed because its CMI components are all significant in the MI identity

$$\hat{I}(S; \{D, A, H\}) = \hat{I}(S; D|\{A, H\}) \ (= 22.368, df = 4, p < 0.001)$$
$$+ \hat{I}(S; H|A) \ (= 71.886, df = 2, p < 0.001)$$
$$+ \hat{I}(S; A) \ (= 88.586, df = 1, p < 0.001). \qquad (10)$$

Further, $\widehat{Int}(S; D; \{A, H\})$ (= 19.690, $df = 3$, $p < 0.001$) and $\widehat{Int}(S; A; H)$ (= 13.543, $df = 1$, $p < 0.001$) are significant by the two-step LR tests.

By deleting insignificant CMI terms (those with asterisks) in (7) to (10), a summary analysis of (6) gives a tentative log-linear model

$$LLM_1 = \{ACS, GMS, ADG, ADHS\}$$

with residual deviance 92.259 ($df = 99$, $p = .671$). Specifically, a more concise model is

$$\hat{I}(C; M; G; S; D; H; A) \cong \hat{I}(C; A|S) + \hat{I}(C; S) + \hat{I}(M; S|G) + \hat{I}(M; G)$$
$$+ \hat{I}(G; D|A) + \hat{I}(G; A) + \hat{I}(S; A) + \hat{I}(S; H|A)$$
$$+ \hat{I}(S; D|\{A, H\}). \qquad (11)$$

At this stage, it is crucial to inspect in detail. It is seen from (8) and (11) that $\widehat{Int}(M; G; S)$ (= 7.22, $df = 1$, $p = .007$) is significant within the CMI $\hat{I}(M; S|G)$ (= 16.935, $df = 2$, $p < 0.001$), and it remains to identify either a single significant GMS term or two significant two-way terms {GM, MS} in $LLM_1$. Meanwhile, a few terms in (11) notably have two letters in common, say, AD, AG, AH, among which, some will be found to be dispensable such that a more parsimonious model than $LLM_1$ is obtained. It follows that a potentially valid model is

$$LLM_2 = \{ACS, ADG, GMS, AHS, DHS\},$$

with residual deviance 116.137 ($df = 102$, $p = .160$). Here, if the common two-way AS effect in {ACS, AHS} is partially excluded from $LLM_2$, say, the three-way ACS is replaced by {AC, CS}, then a parsimonious log-linear model can be derived as

$$\text{LLM}_3 = \{AC, CS, ADG, GMS, AHS, DHS\}, \tag{12}$$

with the residual deviance 124.370 ($df = 103$, $p = .075$) and estimated AIC = 176.46. Since the three-way ADG is already included in the models $\text{LLM}_1$ and $\text{LLM}_2$, the choice of DHS rather than ADS will give more information in $\text{LLM}_3$, although both three-way effects are included in the last summand on the right-hand side of (11). In fact, if DHS is replaced by ADS in (12), it would yield a lack of fit with residual deviance 198.502 ($df = 104$, $p < 0.001$), due to missing the DH effect, $\hat{I}(D; H) = 144.473$ ($df = 1$, $p < 0.001$). Alternately, if the interaction GMS is replaced by the set {GM, MS} in $\text{LLM}_2$, then the most parsimonious model

$$\text{LLM}_4 = \{ACS, ADG, AHS, DHS, GM, MS\} \tag{13}$$

can be derived with the deviance 125.493 ($df = 104$, $p = .074$) and estimated AIC = 175.44. In terms of the AIC estimates, it is clear that both models (12) and (13) are equally effective. The degrees of freedom between models (12) and (13) differ by one, because the AS effect is repeated in {ACS, AHS} in (13). When both ACS and GMS are replaced by {AS, CS} and {GM, MS}, respectively, the resulting model is lack-of-fit with deviance 133.727 ($df = 105$, $p = .031$). In summary, the above discussion concludes that the proposed MI analysis of log-linear models yields two acceptable parsimonious models $\text{LLM}_3$ and $\text{LLM}_4$ for the 7-way contingency table of the ischemic stroke data.

## 4. The MI Analysis of Logit Models

It is well known that a logit model is equivalently given by a valid log-linear model. Thus,

it is expected that parsimonious logit models may be derived from the MI analysis similar to those in the log-linear modelling. Suppose that the target of interest is the ischemic stroke status, defined to be S = 1 for case, and S = 0 for control. The $LLM_4$ model (13) suggests that the association between G (gender) and S need not be significant, whereas $LLM_3$ (12) indicates that the interaction GM is significantly associated with S. Therefore, the following identity can be useful in decomposing the MI between the target and the six risk factors, that is,

$$\hat{I}(\{G, M, H, D, C, A\}; S) = \hat{I}(S; G|\{M, H, D, C, A\})^* (= 28.837, df = 32, p = .627)$$
$$+ \hat{I}(S; M|\{H, D, C, A\}) (= 26.110, df = 16, p = .052)$$
$$+ \hat{I}(S; C|\{H, A, D\})$$
$$+ \hat{I}(S; \{H, A, D\}). \qquad (14)$$

In accordance with equation (10), the last summand of (14) yields significant association between S and the factors age, diabetes mellitus and hypertension, that is, {A, D, H}, an association which has been widely studied, for instance, in Movahed et al. (2010) and Sowers (2013). The first summand in (14) shows that there is no significant effect of G on S, conditional on the other risk factors. The second summand indicates that after removing the factor G, the factor M (smoking) is marginally significant, conditional on the remaining four factors. In view of the third summand, it is worth noting that while this study is mainly interested in the calcification burden in the MCI territory, a clinical question would be whether smoking is related to the ischemic stroke among the risk factors. Indeed, by equation (4), the second summand of (14) can be expressed as the following sum:

$$\hat{I}(S; M|\{H, D, C, A\}) = \widehat{Int}(S; M|\{H, D, C, A\})\ (15.495,\ df = 15,\ p = .416)$$
$$+ \widehat{Par}(S; M|\{H, D, C, A\})\ (10.615,\ df = 1,\ p = .001), \quad (15)$$

which confirms a significant marginal effect of M on the target S, without significant interaction effect between S and M across all levels of {H, D, C, A}. Now, the MI between S and the five risk factors {C, M, H, A, D} can be decomposed by the rule of yielding the least significant higher-order interaction effects, as given in equation (16) and listed in Table 1 below. That is,

$$\hat{I}(S; \{C; M; H; A; D\}) = \hat{I}(S; D) + \widehat{Par}(S; A|D) + \widehat{Int}(S; D; A)$$
$$+ \widehat{Par}(S; H|\{D, A\}) + \widehat{Int}(S; H; \{D, A\})$$
$$+ \widehat{Par}(S; C|\{H, D, A\}) + \widehat{Int}(S; C; \{H, D, A\})^*$$
$$+ \widehat{Par}(S; M|\{C, H, D, A\}) + \widehat{Int}(S; M; \{C, H, D, A\})^*. \quad (16)$$

Table 1. Partitioned CMI terms in the MI identity (16).

| Orthogonal Components | Conditional Mutual Information | | | Interaction | | | Partial Association | | |
|---|---|---|---|---|---|---|---|---|---|
| | LR | df | p | LR | df | p | LR | df | p |
| $\hat{I}(S; M|\{C, H, D, A\})$ | 26.110 | 16 | < 0.001 | *15.495* | *15* | *0.416* | 10.615 | 1 | 0.001 |
| $\hat{I}(S; C|\{H, D, A\})$ | 78.153 | 8 | < 0.001 | *11.963* | *7* | *0.102* | 66.190 | 1 | < 0.001 |
| $\hat{I}(S; H|\{D, A\})$ | 55.444 | 4 | < 0.001 | 12.257 | 3 | 0.007 | 43.187 | 1 | < 0.001 |
| $\hat{I}(S; A|D)$ | 103.314 | 2 | < 0.001 | 27.840 | 1 | < 0.001 | 75.474 | 1 | < 0.001 |
| $\hat{I}(S; D)$ | 24.083 | 1 | < 0.001 | | | | | | |

There are five significant partial association terms and two significant {AD, ADH} interaction terms in the MI identity (16) and Table 1. By possible redundancy of the highest-order SADH interaction, it is commonly suggested that a parsimonious logit model, coined the MI logit model, can be estimated as

$$\text{logit}\left[f(S\,|\,C, M, H, A, D)\right]$$
$$= -3.584 + 1.653D + 1.659A - 1.003DA + 1.689H - 0.864AH$$
$$- 0.763DH + 0.495M + 2.119C, \qquad (17)$$

which gives the residual deviance 26.651 ($df = 23$, $p = 0.271$). By using the same set of five predictors {C, M, H, A, D}, it is remarkable that the minimum AIC model can be easily found among a few neighbors to the MI model (17). In the search, it is convenient to include a few lower-order interaction effects such as {MA, MD, MH} so as to increase the acquired model log-likelihood, in addition to the five main effects and three interaction effects of model (17). The minimum AIC model is confirmed by computing just a few AIC estimates (cf. *SAS CATMOD or SPSS logistic procedure*) as

$$\text{logit}\left[f(S\,|\,C, M, H, A, D)\right]$$
$$= -3.824 + 1.895D + 1.895A - 1.130DA + 1.664H - 0.749DH - 0.841AH$$
$$+ 1.180M - 0.652MA - 0.663MD + 2.083C. \qquad (18)$$

Model (18) consists of significant parameter estimates, it yields the log-likelihood -49.324, the minimum AIC estimate 120.649 and the residual deviance 18.973 ($df = 21$, $p = 0.587$). In contrast, the MI model (17) yields a slightly larger AIC estimate 124.327 but slightly smaller log-likelihood -53.163. With greater model log-likelihood, the AIC (18) acquires better prediction accuracy over the MI (17) by using extra interaction effects {MA, MD}, but loses the ability of interpreting or predicting S by the factor M. Indeed, with the risk factor M, the parameter estimate 0.495M in the MI model (17) is replaced by 1.180M in the AIC model (18), and the latter estimate is clearly farther away from the logarithmic odds ratio estimate 0.489 in the raw data.

It is well known that the minimum AIC model (18) gives the best prediction accuracy which is in principle equivalent to the model selection based on cross-validation (Akaike, 1974; Burnham and Anderson, 2004; Fahrmeir and Tutz, 2013; Kateri, 2014; Stone, 1974). For the present empirical study, it is generally expected that better results of testing for model fit may be obtained with the AIC model (18) because it has two more parameter estimates than the parsimonious MI model (17). The validity of these models can be examined through testing goodness-of-fit using residual deviance statistics with simulated samples under each assumed model, and under different sampling conditions of the raw data. A simulation study of 10,000 replicates of various sample sizes is conducted to compare the validity of these two models by goodness-of-fit test under three sampling designs. The first design assumes sampling under the MI model (17) or the AIC model (18). The second design assumes sampling the empirical multinomial distribution of the raw data with replacement, which is regarded as a restrictive design not generally useful. The third design assumes sampling random subsets of the raw data without replacement. Simulation results of testing for model fit against the MI and AIC models under the assumed sampling designs are reported in Table 2 using two sample sizes, 800 and 1000.

Table 2. Proportions of accepting model (17) or (18) under sampling designs

| Tested models /sample size | True MI model (17) | True AIC model (18) | Raw data multinomial distribution | Raw data random subsets |
|---|---|---|---|---|
| MI (17) / 800 | .9959 | .9780 | .8733 | .9827 |
| AIC (18) / 800 | .9961 | .9978 | .9487 | .9961 |
| MI (17) / 1000 | .9954 | .9639 | .7867 | .9804 |
| AIC (18) / 1000 | .9959 | .9955 | .9040 | .9971 |

It is not surprising that higher acceptance rates of the AIC model than the MI model are obtained under each simulation condition, when more parameter estimates of the same predictors are used in the larger AIC model as compared with the MI model. This may not always hold valid in other empirical studies under the assumption of the true MI model.

## 5. Discussion

In this study, we demonstrate the constructive analysis of log-linear and logit modeling using the geometry of the mutual information defined with the multivariate multinomial distribution. The proposed analysis is illustrated using a thorough empirical study of the ischemic stroke contingency data table. It is essential that the CMI analysis is able to identify the main-effect predictors and their indispensable interaction effects such that the acquired log-linear and logit models are undoubtedly most parsimonious. For a finite dimensional contingency table, the conventional approach to log-linear modelling begins with inspecting two-way association effects and successive testing for higher-order interaction effects. The proposed geometric analysis develops a backward selection scheme by deleting dispensable higher-order interaction effects through the CMI analysis. As a counterpart to log-linear modelling, the MI analysis naturally constructs the information approach to logit modelling for the same empirical study. The acquired MI logit model is most parsimonious, which usually differs from the minimum AIC model when using the same set of five or more predictors. In the current stroke data analysis, it is found that the AIC model with two more interaction parameters could acquire higher proportions of accepting goodness-of-fit test than the MI model does under various sampling designs. This is acquired by the AIC at the cost of losing the data interpretation through a predictor, i.e., the risk factor "smoking" in this study.

It is well known that in the modern analysis of contingency tables, standard methods of variable selection and model selection with log-linear models and logit models are often discussed using AIC, BIC and other penalized criteria. These methods are unable to straightforwardly identify the necessary parameters in a desired model, unlike the constructive schemes developed with the proposed MI analysis. When both categorical and continuous variables are present in the data, it is recommended that multivariate multinomial distributions can be employed to characterize the marginal distributions of the continuous variables by discrete approximations. That is, discretization of the continuous variables as multivariate histograms can be used to describe the MI of all the variables such that the proposed schemes of the MI analysis of log-linear and logit modeling are able to provide an initial analysis by integrating both continuous and discrete variables in the data. It is thus expected that the proposed MI analysis can be applicable to GLMs through discretization of the multivariate sampling distributions, as will be examined in a future study.

Acknowledgment. This study was supported by the grants MOST-105-2410-H-001-036 and MOST-105-2811-H-001-021.